\documentclass{llncs}
\usepackage[english]{babel}  
\usepackage{hyperref,graphicx,alltt,xcolor,float}
\usepackage{color}
\usepackage{url}

\begin{document}

\title{Towards an Integrated Penetration Testing Environment for the CAN Protocol}

\author{Giampaolo Bella\inst{1} \and Pietro Biondi\inst{2}}

\institute{Dipartimento di Matematica e Informatica, Universit\`a di Catania, Italy
\email{giamp@dmi.unict.it}
\and
Dipartimento di Matematica e Informatica, Universit\`a di Catania, Italy
\email{pietro.biondi94@gmail.com}}

\maketitle

\begin{abstract}
The Controller Area Network (CAN) is the most common protocol interconnecting the various control units of modern cars. Its vulnerabilities are somewhat known but we argue they are not yet fully explored --- although the protocol is obviously not secure by design, it remains to be thoroughly assessed how and to what extent it can be maliciously exploited. This manuscript describes the early steps towards a larger goal, that of integrating the various CAN pentesting activities together and carry them out holistically within an established pentesting environment such as the Metasploit Framework. In particular, we shall see how to build an exploit that upsets a simulated tachymeter running on a minimal Linux machine. While both portions are freely available from the authors' Github shares, the exploit is currently subject to a Metasploit pull request.

\end{abstract}

\section{Introduction}
When traditional security vulnerabilities are ported to and exploited in the automotive environment, they bear a clear potential to compromise passengers' safety; therefore, 
modern automotive technology intertwines security and safety requirements tightly and is, as such, worth of considerable attention.

The most widespread protocol that connects the various control units found in modern cards is the \textit{Controller Area Network} (CAN) protocol, which appears to be highly vulnerable at least because it is not meant to be secure by design.
However, a full understanding of the technicalities of this protocol and  its vulnerabilities is still out of reach for a variety of reasons, such as the scattered and mostly unofficial documentation, the large customisation operated by each car manufacturer, and the lack of an integrated pentesting environment to investigate vulnerabilities and attempt to exploit them.

The CAN protocol is standardised by ISO 11898-1:2015 \cite{isocan}, though without any reference to security issues,
perhaps in the assumption that a car forms a secluded, protected network environment. 
The question arises on what would happen should the CAN bus be used in an unprotected environment instead, namely with potentially malicious activity running through its wires.

The security protocol literature is full of examples of protocols or systems devised to stand a threat model but, implicitly, not a stronger one. The best known example is the public-key Needham-Schroeder protocol \cite{ns} and Lowe's attack on it \cite{ns-lowe}. The attack originated from an initiator who started with the attacker, but Needham commented that their protocol was not meant to withstand insider threats \cite{needhamrefusal}.

It is clear that the mentioned assumption that a car hosts an isolated network falters at present, as the control and infotainment systems are often combined and interfaced with the external world through the Internet. For example, one can check the pressure of the tyres or switch on the engine \textit{remotely} through an app running on their smartphone. In this socio-technical context, attacks are starting to appear. For example, a Jeep Cherokee was hacked via bluetooth, with the attacker being able to remotely operate the brake system and the steering wheel \cite{remoteattack}; a Toyota Prius was hacked via the JTAG port and the entire CAN bus traffic hijacked, even letting the attacker flash the firmware of a control unit\cite{caninject}.
%

This manuscript aims at contributing to the definition of an integrated environment to conduct simulation experiments on the CAN bus. In short, it contributes a lightweight Linux machine running a tachymeter simulator, and the machine itself is made vulnerable to simulate an in-vehicle network that is accessed, for example, via bluetooth or the JTAG port. It also brings a Metasploit post exploitation module as a Ruby file {\tt crazytachymeter.rb}, which causes the tachymeter to jump without continuity to random speeds, also outside the programmed range.

More precisely, these contributions derive from the following research methodology. The Instrument Cluster Simulator (ICSim) \cite{carbook} is installed on the Linux machine and operated; the generated traffic is sniffed and interpreted using Kayak \cite{kayak}. Potentially meaningful values for the data frames are then conjectured and verified manually. After that, the findings are used to program the script in the Ruby file, which is then tested in the Metasploit framework \cite{metasploit} and, at the same time, submitted for consideration and inclusion in future releases of the framework.


The structure of the paper follows the methodology just outlined. A primer on the CAN bus supports the developments (\S\ref{sec:can}). The tachimeter simulator is introduced (\S\ref{sec:icsim}), along with Kayak (\S\ref{sec:kayak}). Then, the exploitation of the conjectured vulnerabilities is carried out manually (\S\ref{sec:exploit}). Finally, the integrated pentesing environment is described and made publicly available (\S\ref{sec:penenv}), and the manuscript draws its conclusions
(\S\ref{sec:concl}).

\section{A primer on the Controller Area Network protocol}\label{sec:can}
Modern cars are full of electronics. Components such as airbags, power doors, electric mirrors need to be interconnected and communicate among each other to ensure the smooth and synergistic functioning of all. This was the aim for the inception of the Can protocol \cite{summacan}, also known as \textit{CAN bus}, which dates back to 1983 at Bosch.

The CAN bus is conceptually simple from a hardware standpoint, as it consits of two wires, \textit{CAN high} (CANH) and \textit{CAN low} (CANL).
It is equally simple from a software standpoint, as it sees a body computer read environmental data such as brake pedal pressure and air conditioning temperature, and then send appropriate commands on the CAN bus to the dedicated control units; this is done in a multicast style, namely commands are sent to all but filtered by the intended recipients.
The body computer must use two CAN wires ensuring fault tolerance, which can be done by means of differential signaling: to send a signal, it must raise the voltage on one line and equally drop the other line.

As mentioned above, the CAN frames are standardised as ISO 11898-1:2015 \cite{isocan} to contain various fields. These include an Arbitration field carrying the frame ID, also used for arbitration, a Control field for control signals and a Data field for the payload. More precisely, the fields are:
\begin{itemize}
	\item Start Of Frame: a dominant bit indicating the beginning of a frame;
	\item Data: up to 8 bytes of data;
	\item Arbitration field: 11 bits identifying the intended recipient from the frame; one bit is the Remote Transmission Request (RTR) bit, which is low for a Data Frame and 1 for a Remote Frame (one whose Data Field is empty);
	\item Control: 6 bits, with 4 bits called Data Length Code (DLC) indicating the length of the Data Field and 2 reserved bits;
	\item CRC: 15 bits for a cyclic redundancy check code and a recessive bit as a delimiter;
   \item Ack: 2 bits, with the first one being recessive, hence overwritten with a dominant bit by every node that receives it, and the second bit working as a delimiter;
   \item End Of Frame: 7 recessive bits.
\end{itemize}

However, it must be noted that car manufacturers interpret the standard freely, for example using padding at will, raising a general issue of how to interpret CAN bus traffic on various cars.

\section{Instrument Cluster Simulator}\label{sec:icsim}
\begin{figure}[H]
	\centering\includegraphics[width=0.9\linewidth]{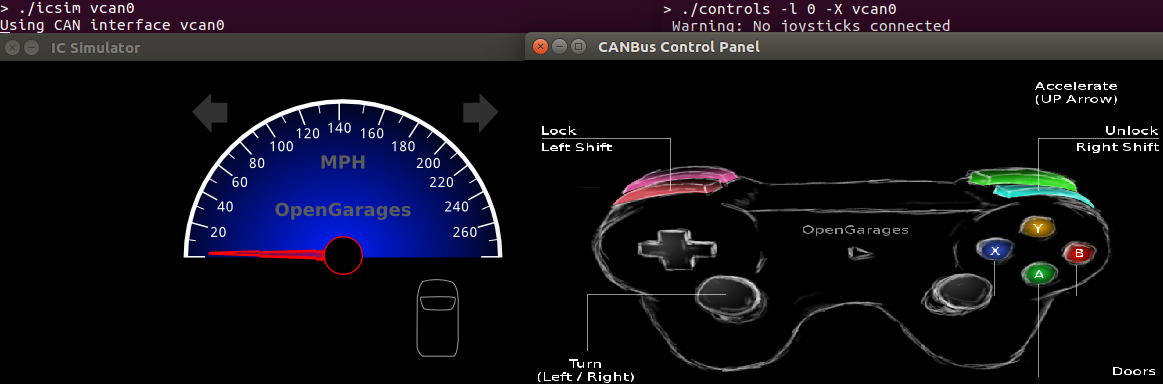}
	\caption{\label{fig:icsim}The control panel of the Instrument Cluster Simulator}
\end{figure}

Instrument Cluster Simulator (ICSim) \cite{carbook} is a simulator for some of the main car functions, namely blinkers, power doors and tachymeter, operated through the CAN bus. It runs on Linux following the setup of a virtual CAN interface through the following simple commands:
\begin{alltt}
\ \ \ \	sudo apt install can-utils	
\ \ \ \	sudo modprobe can
\ \ \ \	sudo modprobe vcan
\ \ \ \	sudo ip link add dev vcan0 type vcan
\ \ \ \	sudo ip link set up vcan0
\end{alltt}

The control panel of the simulator is in Figure \ref{fig:icsim}. Accelerations can be triggered by pushing the Accelerate button; everytime the button is pressed, a specific frame is sent to the control unit of the tachymeter. The simulator supports speed up to 100MPH, as we realise that pressing the button additionally when speed is at 100MPH produces no effect.

\section{Kayak}\label{sec:kayak}
Kayak is an application that sniffs CAN bus traffic \cite{kayak}. It is written in Java, hence easily portable, and features an intuitive user interface. We set out to interpret the CAN traffic generated through ICSim using Kayak. The aim is to understand which frame IDs are associated to which device of the car, whether blinkers, doors or tachymeter, and what values the Data field accepts for each device.

\begin{figure}[h]
	\centering\includegraphics[width=0.95\linewidth]{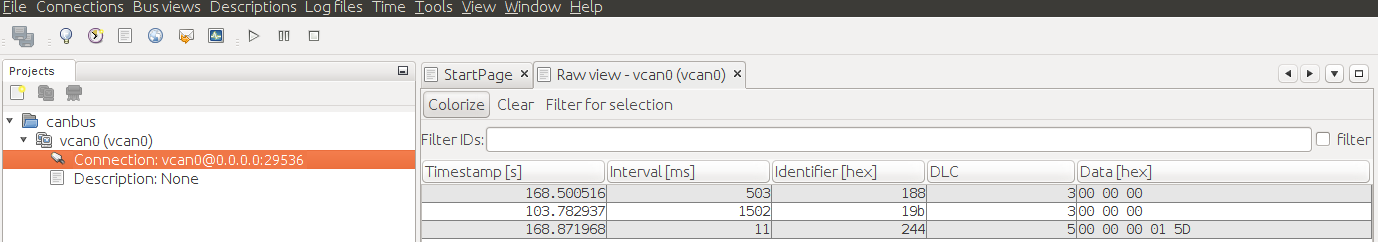}
	\caption{\label{fig:kayak}Packet sniffing on ICSim traffic through Kayak}
\end{figure}

The graphical interface of Kayak can be seen from Figure \ref{fig:kayak}. It features the following fields:
\begin{itemize}
\item Timestamp (seconds): when the frame is intercepted;
\item Interval (milliseconds): interval between two statistics transmissions;
\item Identifier (hexadecimal): corresponding to the frame ID;
\item Data Length Code: length of the Data field
\item Data (hexadecimal): actual payload.
\end{itemize}

Various interactions with the control panel of ICSim can be orchestrated and closely observed through the sniffer. It was easy to derive the information summarised in Table \ref{table:sniffing}. The Table shows the IDs associated to which device, in particular with 5 bytes devoted to data for the tachymeter. The third column emphasises in red positions of the hexadecimal numbers that are observed to change for each device, for example the last four in case of frames for the tachymeter. 

\begin{table}
\begin{center}
	\begin{tabular}{|c|c|c|c|c|}
		\hline
		ID & DLC & Data & Device & Values \\ 
		\hline \hline
		19b & 3 & 00 00 0\textbf{\textcolor{red}{0}} & doors & 1/2/4/8 \\ 
		\hline 
		188 & 3 & 0\textbf{\textcolor{red}{0}} 00 00 & blinkers & 1/2 \\ 
		\hline 
		244 & 5 & 00 00 00 \textbf{\textcolor{red}{00 00}} & tachymeter & 00 00 \ldots 01 5D\\ 
		\hline 
	\end{tabular} 
\end{center}
\caption{Interpretation of ICSim traffic through Kayak}
\label{table:sniffing}
\end{table}
It can also be seen that there are four possible values transmitted to operate the doors, one per door and, likewise, two for the blinkers. As for the tachymeter, the allowed speed range of 0-100MPH was found to be triggered by hexadecimal data values ranging from 00 00 to 01 5D.

\section{Pentesting the Tachymeter of ICSim}\label{sec:exploit}
%
%
We carry out a few penetration testing experiments on the tachymeter of ICSim by conjecturing that it would accept arbitrary frames. The suite of commands previously installed through {\tt can-utils} (\S\ref{sec:icsim}) turns out useful to explore the conjectures. One of the first meaningful conjectures is to send the highest possible hexadecimal value 99 99 to the tachymeter and observe whether the highest possible speed is reached. This can be verified by sending command:
\begin{alltt}
\ \ \ \	cansend vcan0 244#0000009999
\end{alltt}
The tachymeter reacts by reaching its top speed, of 240MPH. It is then possible to send arbitrary values and observe the corresponding speeds each time.

A similar conjecture is about exploiting the blinkers. Coherently with the lessons learned using Kayak (Table\ref{table:sniffing}, \S\ref{sec:kayak}), we can try out command:
\begin{alltt}
\ \ \ \	cansend vcan0 188#030000
\end{alltt}
As a result, both blinkers would turn on.

Even though in a simulated environment, these experiments reconfirm what we already knew: no security measures whatsoever are implemented, not even an out-of-range check.

\section{Towards the Integrated Pentesting Environment}\label{sec:penenv}
Our aim of an integrated pentesting environment can be pursued by taking two main steps. The first step is the preparation of a machine to simulate the victim system, and this requires a few sub-steps in turn. 

We build a machine running Bodhi Linux, a minimal distribution based on Ubuntu 16.04 LTS, and implement a vulnerable python server to run on it. The core of the server receives data off a socket and executes it, as stated by line {\tt data=c.recv(5120) subprocess.call(data,shell=True)}. This is meant to simulate a successful malicious access to the network laid within the car.
We then install ICSim on the machine, and our simulated victim is ready. It can be downloaded from our GitHub share \cite{biondirepo}.

The second step is the automation of the pentesting experiments outlined above (\S\ref{sec:exploit}) through an actual exploit executable on the Metasploit Framework \cite{metasploit}. We wrote such an exploit in Ruby and submitted file {\tt crazytachymeter.rb} to Metasploit for consideration; as a possible location to host the exploit, our pull request \cite{pullrequest} pinpointed path {\tt modules/exploits/unix/misc/}. 

The pull request is currently ongoing, and the interaction has been fruitful for us. We have been advised to treat our program as a post-exploitation script, rather than as an actual exploit, due to the fact that it targets a tailored machine and follows a successful malicious access to have occurred beforehand. The new path therefore is {\tt modules/post/hardware/automotive/}. This already contains five scripts, which, incidentally, are worth of consideration here:
\begin{itemize}
\item {\tt canprobe.rb} allows the analyst to 
scan for given frame IDs and set their data fields;
\item {\tt getinfo.rb} returns engine and vehicle information;
\item {\tt identifymodules.rb} searches for devices responding to Diagnostic Session Control (DSC) queries;
\item {\tt malibu\_overheat.rb} controls the temperature gauge of a 2006 Chevrolet Malibu;
\item {\tt pdt.rb}: discovers the Pyrotechnic Control Units (PCU) units and sets them ready to be deployed.
\end{itemize}	
%
%
%

\begin{figure}[h]
	\centering
	\includegraphics[scale=.4]{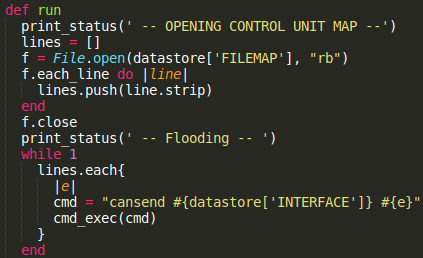}
	\caption{\label{fig:exploit}Our script: the core}
\end{figure}

The gist of our script is pictured in Figure \ref{fig:exploit}. It can be seen that it takes as a parameter a file containing a specific mapping of the frames; the mapping represents the interpretation of the frame by a control unit, and therefore, the script is general and not bound to a single mapping. The script then loops forever while it sends frames to the virtual CAN interface. Also the script is available from GitHub \cite{biondirepo}.

\begin{figure}[h]
	\centering
	\includegraphics[width=0.95\linewidth]{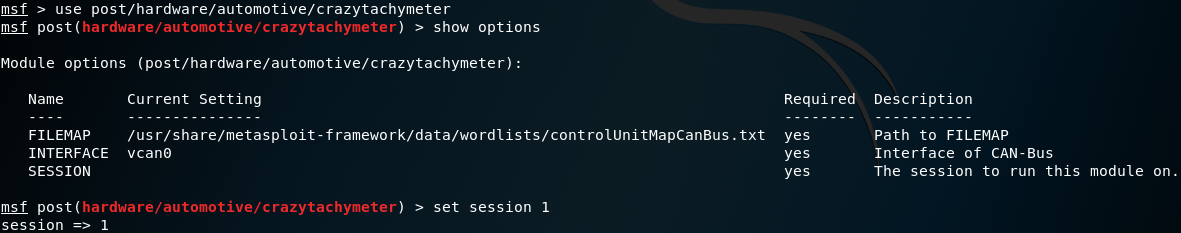}
	\caption{\label{fig:options}Our script: the options}
\end{figure}

While the pull request is being processed, our script can be manually downloaded and run over Metasploit, as shown in Figure \ref{fig:options}. It exposes the three intended module options: {\tt FILEMAP}, {\tt INTERFACE} and {\tt SESSION}.

\begin{figure}[h]
	\centering
	\includegraphics[scale=.35]{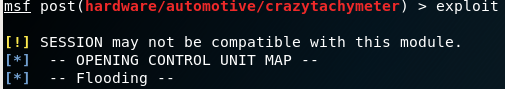}
	\caption{\label{fig:exploit_run}Our script: the output}
\end{figure}

Once the options are set, the script can be run. Figure \ref{fig:exploit_run} shows that it returns the {\tt Flooding} message, which means that the tachymeter is being flooded with frames corresponding to various speeds.

\begin{figure}[h]
	\centering
	\includegraphics[scale=.45]{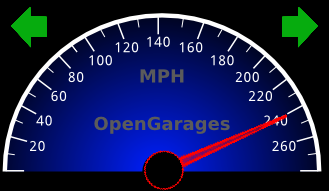}
	\caption{\label{fig:maxspeed}Our script: a snapshot of its consequences}
\end{figure}

The visible outcome of running the script is that the tachymeter goes crazy by jumping up or down without continuity over the range 0-240MPH. Of course, this cannot be portrayed in a picture, and Figure \ref{fig:maxspeed} provides the single screenshot of the top speed.

\section{Conclusions}\label{sec:concl}
Investigations into the cybersecurity limitations of the CAN protocol have only just began, with the most significant findings dating back to a bunch of years ago. The fact that the protocol never meant to be secure by design cannot decrease our surprise at how easy it turns out to be to exploit nodes that run it, as shown above.

This paper described how to send a tachymeter crazy. Starting from a tachy\-meter simulator, it was possible to decode the frame data values that would trigger specific events, and hence to try out additional data values at will. Doing so revealed the possibility of inducing anomalous scenarios, such as the blinkers turning on on both sides and the tachymeter jumping from one speed to another --- and beyond the programmed range of 0-100MPH. The tachimeter runs on a Bodhi Linux machine made vulnerable, and the attacks are implemented as Metasploit post-exploitation modules.

Although our experiments only took place on a simulated environment, they highlight that no security measure is in place against malicious activity on the CAN bus. This finding may not be surprising by itself. However, it required the gathering of a number of tools and their combined use, and these activities were more time consuming than expected.

Where do we go from here? On one hand, it is all the more clear that the CAN protocol ought to be amended to incorporate even simple security measures that would control and qualify the frames. Such measures are currently being studied, and their deployment would subvert the finding that, once an attacker penetrates the in-vehicle network, they can then command and control all nodes on the CAN bus.

On the other hand, we realise that, if it is arguably complicated and somewhat expensive to setup a real laboratory to experiment on the CAN bus, it  should really be made simple to conduct the experiments in a simulated environment. This manuscript contributed in such a direction by showing that a Linux machine (or a network of machines) can be tailored to simulate the nodes and network laid within a car. It then suggested to write the potential attacks in Ruby so that they could be simulated using Metasploit. If the machine-exploit pair is easy do download (such as ours \cite{biondirepo}), then simulations on the CAN bus could finally fall within effortless reach.

\bibliographystyle{splncs}
\bibliography{automotivebiblio}

\end{document}